%% file: paper.tex
\documentclass{ar-1col}
\usepackage[numbers]{natbib}
\usepackage{url}
\usepackage{xspace}
\usepackage{amsmath}
\usepackage{amssymb}
\usepackage{amsfonts}
\usepackage{upgreek}

\jname{Annual Review of Nuclear and Particle Science}
\jvol{71}
\jyear{2021}
\doi{10.1146/annurev-nucl-110320-051823}
\firstpagenote{Posted with permission from the Annual Review of Nuclear and Particle Science, Volume 71 \copyright\ 2021 by Annual Reviews, \url{http://www.annualreviews.org/}}

\definecolor{greenp1}{rgb}{0, 0.8, 0}
\definecolor{ebdcolor}{rgb}{0.6, 0, 0}
\definecolor{pbdcolor}{rgb}{0, 0.6, 0.6}
\definecolor{mdcolor}{rgb}{0.6, 0, 0.6}
\definecolor{eecolor}{rgb}{0, 0.6, 0}
\definecolor{eftcolor}{rgb}{0.6, 0.6, 0}
\definecolor{ppcolor}{rgb}{0, 0, 0.6}

\def\aprime{\ensuremath{A^{\prime}}\xspace}
\def\ma{\ensuremath{m_{\aprime}}\xspace}
\def\ta{\ensuremath{\tau_{\aprime}}\xspace}
\def\BL {\ensuremath{B\!-\!L}\xspace}

\begin{document}

\markboth{Graham--Hearty--Williams}{Searches for dark photons at accelerators}

\title{Searches for dark photons at accelerators}

\author{Matt Graham,$^1$ Christopher Hearty,$^2$ and Mike Williams$^{3,4}$
\affil{$^1$SLAC National Accelerator Laboratory, Stanford University, Stanford, CA 94309, USA}
\affil{$^2$University of British Columbia / IPP, Vancouver, Canada, V6T 1Z1}
\affil{$^3$Massachusetts Institute of Technology, Cambridge, MA 02139, USA}
\affil{$^4$Monash University, Melbourne, Victoria 3168, Australia}}

\begin{abstract}
Dark matter particles may interact with other dark matter particles via a new force mediated by a {\em dark photon}, \aprime, which would be the dark-sector analog to the ordinary photon of electromagnetism.
The dark photon can obtain a highly suppressed mixing-induced coupling to the electromagnetic current,
providing a portal through which dark photons can interact with ordinary matter.
This review focuses on \aprime scenarios that are potentially accessible to accelerator-based experiments.
We summarize the existing constraints placed by such experiments on dark photons, highlight what could be observed in the near future, and discuss the major experimental challenges that must be overcome to improve sensitivities.
\end{abstract}

\begin{keywords}
dark photons, dark matter, hidden sectors, non-minimal models, accelerators
\end{keywords}
\maketitle

\tableofcontents

\section{INTRODUCTION}
\label{sec:intro}
\input{intro}

\section{PHENOMENOLOGY}
\label{sec:theory}
\input{theory}

\section{SEARCH STRATEGIES}
\label{sec:strategy}
\input{strategy}

\section{SEARCHES FOR VISIBLE DARK PHOTONS}
\label{sec:visible}
\input{visible}

\section{SEARCHES FOR INVISIBLE DARK PHOTONS}
\label{sec:invisible}
\input{invisible}

\section{RICH DARK SECTORS}
\label{sec:rich}
\input{rich}

\section{SUMMARY \& OUTLOOK}
\label{sec:sum}
\input{sum}

\section*{DISCLOSURE STATEMENT}
MW is a member of the LHCb collaboration.
CH is a member of the BaBar and Belle II collaborations.
MG is a member of the BaBar, HPS, LDMX, and DarkQuest collaborations.

\section*{ACKNOWLEDGMENTS}
We thank Gordan Krnjaic for providing several of the exclusion regions in Fig.~\ref{fig:invisible-limits-DM}.
All other constraints were obtained from Ref.~\cite{darkcast}.
The authors were supported by the following grants:
MG by the U.S. Department of Energy under Contract No. DE-AC02-76SF00515,
CH by Natural Sciences and Engineering Research Council of Canada,
and MW by NSF grant PHY-1912836.

\appendix
\section{DESCRIPTION OF EXPERIMENTS}
\label{sec:experiments}
\input{experiments}

\bibliographystyle{ar-style5}
\bibliography{dark-photons-new,dark-photons}

\end{document}

%% file: intro.tex
It would be difficult to overstate the success of the Standard Model (SM) of particle physics; however, the SM cannot be a complete theory of nature since, {\em e.g.}, it cannot explain dark matter.
The existence of dark matter is firmly established due to its gravitational interactions, but little is known about the dynamics within the dark sector itself.
An intriguing possibility is that dark matter particles may interact with other dark matter particles via a new {\em dark force}, similar to the electromagnetic force felt by ordinary matter.
If this is the case, then one expects there to be a {\em dark photon}, \aprime, that mediates this dark force, in analogy to the ordinary photon of electromagnetism.
This exciting possibility has motivated a dedicated worldwide effort to search for dark photons and other dark sector particles (see, {\em e.g.}, Refs.~\cite{Battaglieri:2017aum,Fabbrichesi:2020wbt} for recent reviews).

In the standard dark sector paradigm, no SM particles are charged directly under any dark-sector interactions and {\em vice versa}.
Therefore, the dark photon does not couple directly to SM particles; however,
it can obtain a small coupling to the electromagnetic (EM) current due to {\em kinetic mixing} between the SM hypercharge and \aprime field strength tensors~\cite{Fayet:1980rr,Fayet:1980ad,Okun:1982xi,Galison:1983pa,Holdom:1985ag,Pospelov:2007mp,ArkaniHamed:2008qn,Bjorken:2009mm}.
This mixing-induced coupling, which is suppressed relative to that of the photon by a factor labeled $\varepsilon$,
provides a {\em portal} through which dark photons can interact with SM particles: dark photons can be produced in the lab, and they can decay into visible SM final states---though decays into (nearly) invisible dark-sector final states are expected to be dominant if kinematically allowed.
One striking advantage of producing dark matter in the lab is that it will be relativistic, which leads to accelerator-based experiments having similar sensitivity to most types of dark matter particles. 
This is in stark contrast to direct-detection experiments, which, {\em e.g.}, have much better sensitivity to scalars than fermions. 

The minimal dark-photon model only has 3 unknown parameters:  the strength of the kinetic mixing, $\varepsilon$; the dark photon mass, \ma; and the decay branching fraction of the dark photon into invisible dark-sector final states, which is typically assumed to be either unity or zero (corresponding to whether any invisible dark-sector final states are kinematically allowed or not).
This review focuses on the region of $[\ma,\varepsilon]$ parameter space accessible to accelerator-based experiments, namely $\ma \gtrsim 1$\,MeV and $\varepsilon \gtrsim 10^{-7}$ (see Ref.~\cite{Fabbrichesi:2020wbt} for a summary of non-accelerator-based constraints on dark photons).
In addition, we concentrate on \aprime masses below the electroweak scale, where dark photon phenomenology is markedly different than supersymmetry and other scenarios that extend the SM.

Thus far, no accelerator-based dark-photon searches have found any evidence for a signal.
Therefore, we will focus on summarizing the constraints placed by accelerator-based experiments on both visible and invisible dark photons.
In addition, this review will highlight what could be observed in the near future, while also discussing the major experimental challenges that must be overcome to improve sensitivities.
Finally, other {\em non-minimal} models will be discussed, where the coupling of the dark boson arises from a different mechanism and/or where other dark-sector particles impact the observable dark-boson phenomenology at accelerator-based experiments.

%% file: theory.tex
This section provides an overview of dark photon phenomenology, including both the theoretical and astrophysical motivations for dark photons.
In addition, dark photon production and decay rates are discussed.

\subsection{The Dark-Photon Portal}
\label{sec:aprime-portal}

The minimal dark photon scenario involves a broken $U(1)'$ gauge symmetry in the dark sector whose field strength tensor, $F'_{\mu\nu}$, kinetically mixes with the SM hypercharge field strength tensor, $B_{\mu\nu}$, via the operator $F'_{\mu\nu}B^{\mu\nu}$.
After electroweak symmetry breaking, and with the gauge boson kinetic terms diagonalized, the dark photon obtains a suppressed coupling to the EM current, $J^{\mu}_{\rm EM}$, where the relevant terms in the Lagrangian are
\begin{align}
	\label{eq:LgammaAp}
	\mathcal{L}_{\gamma A^\prime}  \supset&
	- \frac{1}{4}F'_{\mu\nu}F^{\prime\mu\nu}
	+ \frac{1}{2} m^2_{A^\prime} A^{\prime \mu} A^{\prime}_{\mu}
	+ \varepsilon \, e \,  A^\prime_\mu J^\mu_{\rm EM} + \mathcal{L}_{\aprime \chi\chi}\, ,
\end{align}
where $\chi$ denotes the lightest dark-charged particle (presumably the dark matter), and the form of the $\aprime\chi\chi$ interaction is left unspecified ($\chi$ could be a scalar, fermion, {\em etc.}). 
The minimal dark-photon model only has 3 unknown parameters:  \ma, $\varepsilon$, and the $\aprime \to \chi\bar{\chi}$ decay branching fraction, which as discussed above, is taken to be either unity or zero depending on whether any invisible dark-sector final states are kinematically allowed; {\em i.e.}, depending on whether $m_{\aprime} > 2 m_{\chi}$. 
A model-dependent coupling to the weak $Z$ current also exists, 
though this appears at $\mathcal{O}(m^2_{A'}/m^2_Z)$ and is only relevant for $\ma \gtrsim 10$\,GeV.
Furthermore, precision electroweak measurements restrict the expanded mixing parameters to be roughly those of Refs.~\cite{Cassel:2009pu,Cline:2014dwa}, leaving $\varepsilon$ as the only free mixing parameter.

\begin{figure}[t]
  \centering
  \includegraphics[width=0.9\textwidth]{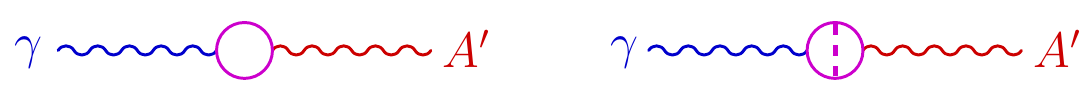}
  \caption{
	Feynman diagrams depicting the generation of kinetic mixing at the (left) 1-loop and (right)~2-loop levels.
  }
  \label{fig:kinmix}
\end{figure}

In principle, the strength of the kinetic mixing is {\em a priori} unknown; however, it is possible to define a target range of $\varepsilon$ to explore by assuming that the mixing arises due to the quantum effects of high-mass particles.
For example, if a heavy particle does exist that carries both hypercharge and dark charge, kinetic mixing could be generated at the one-loop level as shown in Fig.~\ref{fig:kinmix}.
Alternatively, if both sectors are part of a larger Grand Unified Theory (GUT) of nature, then the leading contribution to the mixing arises at the two-loop order.
Fully exploring this {\em few-loop} range of kinetic-mixing strength, which roughly corresponds to $10^{-6} \lesssim \varepsilon \lesssim 10^{-2}$ even if the relevant high-mass particles are at the Planck scale, is an important milestone of dark-sector physics, though values of $\varepsilon$ outside of this range could also arise, {\em e.g.}, if the mixing is nonperturbative.

\subsection{Dark-Photon Production}

Since dark photons couple to SM particles via the ordinary EM current, suppressed by $\varepsilon$, they can be produced anywhere a virtual photon with mass \ma can be produced.
The production mechanisms that have been exploited by accelerator-based dark-photon searches can be categorized as follows:
\begin{list}{\textbullet}{\leftmargin=1em}
\item bremsstrahlung, $eZ\to eZ\aprime$ and $pZ\to pZ\aprime$, where an incident electron or proton radiates a dark photon during an interaction with a fixed nuclear target of charge $Z$;
\item annihilation, $e^+e^-\to\aprime\gamma$, at an $e^+e^-$ collider;
\item Drell-Yan~(DY), $q\bar{q}\to\aprime$, where a quark and anti-quark annihilate into a dark photon, which could occur at a hadron collider or when a proton-beam is incident on a fixed nuclear target;
\item meson decays, {\em e.g.}\ $\pi^0\to\aprime\gamma$ or $\eta\to\aprime\gamma$, for dark photons with $\ma < m_{\pi,\eta}$ at any experiment where mesons are produced at high rates; 
\item and $V\to\aprime$ mixing, where $V=\omega,\rho,\phi$ denotes the QCD vector mesons, which is important for $\mathcal{O}($GeV$)$-scale dark photons.
\end{list}
Proposed future searches largely exploit the same set of production mechanisms (see Sec.~\ref{sec:strategy}).
We note that other production mechanisms, {\em e.g.}\ secondary positrons produced in beam dumps subsequently annihilating~\cite{Marsicano:2018glj}, are possible though not considered in currently published constraints. 

\subsection{Dark-Photon Decays}
\label{sec:aprime-decays}

\begin{figure}[t]
  \centering
  \includegraphics[width=0.7\textwidth]{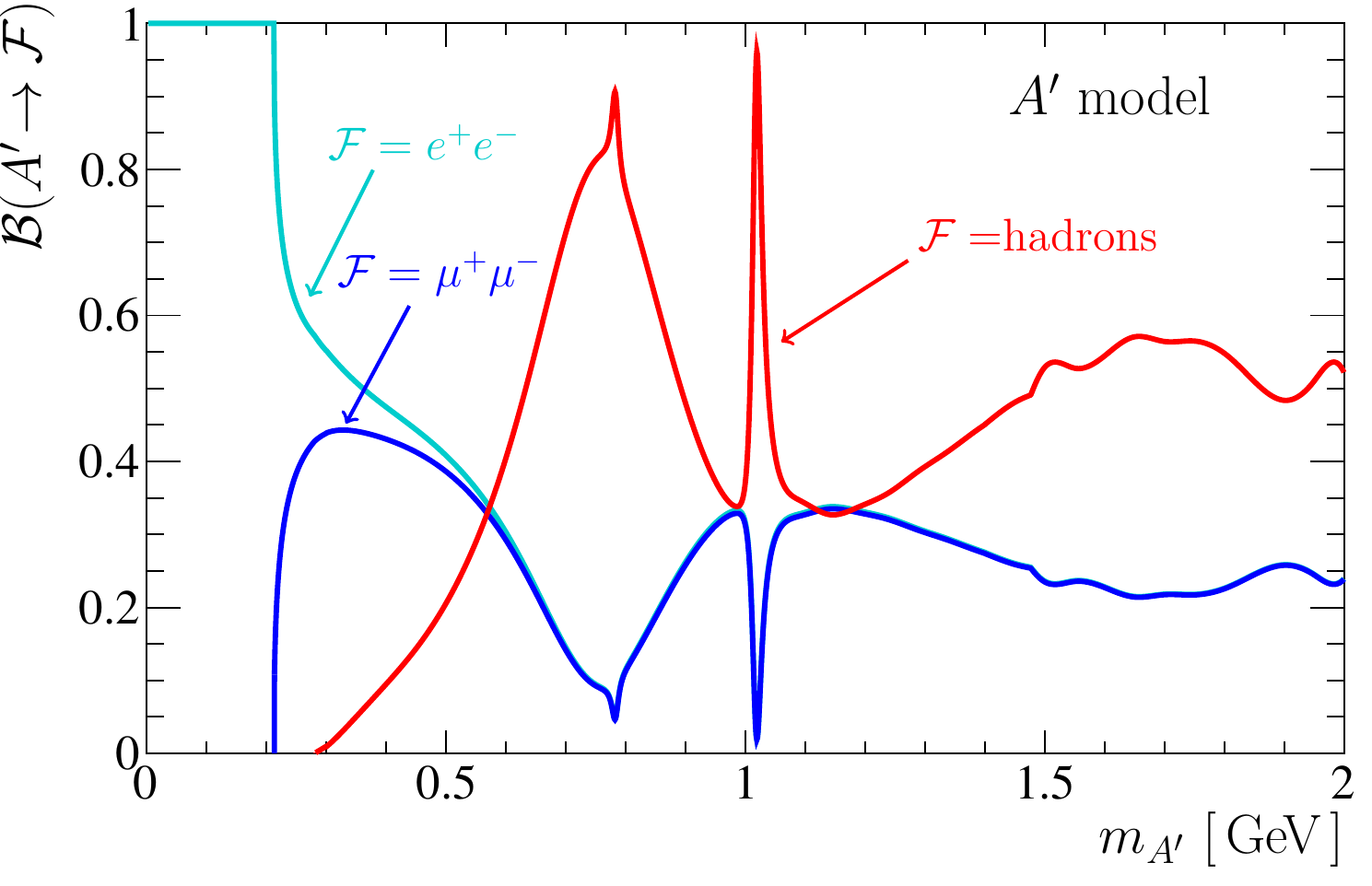}
  \caption{
   From Ref.~\cite{Ilten:2018crw}: dark photon decay branching fractions for the visible dark photon scenario for $\ma < 2$\,GeV. It is straightforward to determine the decay rates into specific hadronic final states by replacing the inclusive hadronic cross section in $\mathcal{R}_\mu$ with the relevant exclusive cross sections.
  }
  \label{fig:aprime_decays}
\end{figure}

Dark photons are expected to decay predominantly into invisible dark-sector final states if kinematically allowed.
If no such decays are allowed, {\em e.g.}\ if $\ma < 2m_{\chi}$, 
then the dark photon will decay into visible SM final states, again due to its suppressed coupling to the EM current.
The partial decay widths of the dark photon into SM leptons are
\begin{align}
	\label{eq:Ap2ll}
	\Gamma_{A'\to \ell^+\ell^-}
=	\tfrac{\varepsilon^2 \alpha_{\rm EM}}{3} m_{A'} \left(  1 +2\tfrac{m^2_\ell}{m^2_{A'}} \right) \sqrt{ 1 - 4\tfrac{m^2_\ell}{m^2_{A'}} } \, ,
\end{align}
where $\ell=e,\mu,\tau$ and, of course, only decays into leptons for which $m_{A'} > 2m_{\ell}$ are allowed.
Decays into hadronic final states cannot be calculated perturbatively for GeV-scale dark-photon masses; however, since
the dark photon couples to $J^\mu_{\rm EM}$, its hadronic decay width can be extracted from the experimentally measured value of $\mathcal{R}_\mu \equiv \sigma_{e^+e^-\to {\rm hadrons}} / \sigma_{e^+e^-\to \mu^+\mu^-}$:
\begin{align}
	\label{eq:Ap2had}
	\Gamma_{A' \to {\rm hadrons} }
= 	\Gamma_{A'\to \mu^+\mu^-} \mathcal{R}_{\mu} (m^2_{A'}) \, .
\end{align}
Equation~\eqref{eq:Ap2had} automatically accounts for mixing with the QCD vector mesons, the $\rho$, $\omega$, and $\phi$, along with all other nonperturbative QCD effects.
Figure~\ref{fig:aprime_decays} shows the branching fractions into $e^+e^-$, $\mu^+\mu^-$, and to all hadronic final states for $\ma < 2$\,GeV.
At higher masses, $\aprime \to q\bar{q}$ decays are easily calculable perturbatively, making determining the branching fractions straightforward.
Finally, the dark photon lifetime, \ta, which is just $\Gamma_{\aprime}^{-1}$, clearly scales as $[\varepsilon^2 \ma]^{-1}$, {\em i.e.}\ the longevity of the dark photon increases as its mass and kinetic-mixing strength decrease.

\subsection{Thermal Dark Matter Targets}

Assuming that dark matter couples strongly enough to ordinary matter that thermal equilibrium was reached in the early universe, there must be some process that subsequently depleted the dark matter abundance. 
If $m_{\chi} > m_{\aprime}$, the purely dark-sector annihilation process $\chi\bar{\chi} \to \aprime \aprime$ is kinematically allowed even at low temperatures. 
Since this process does not depend on $\varepsilon$, we cannot define a precise thermal target if this annihilation scenario is dominant. 
However, if $m_{\chi} < m_{\aprime}$, 
dark matter annihilation is via $\chi\bar{\chi} \to A^{\prime*} \to f\bar{f}$, where $f$ denotes a charged SM fermion. 
This process does depend on $\varepsilon$, which must be large enough, and $m_{\aprime}$ small enough, to achieve the required thermal relic cross-section. 
It is standard to define the dimensionless interaction strength $y$ such that
\begin{align}
\langle \sigma v \rangle \propto \frac{\varepsilon^2 \alpha_D m_{\chi}^2}{m_{\aprime}^4} \equiv \frac{y}{m_{\chi}^2},
\end{align}
where $\alpha_D$ is dark-sector fine-structure constant. 
Using this convention, for any choice of dark matter mass a unique value of $y$ is compatible with thermal freeze-out independently of the specific values of the other parameters, in the limit $m_\chi \ll m_{\aprime}$. 
In addition, for any choice of $\alpha_D$ and the ratio $m_{\aprime}/m_{\chi}$, we can then determine the smallest value of $\varepsilon$ that is consistent with thermal equilibrium in the early universe. 
For Dirac fermions, CMB data from Planck~\cite{Ade:2015xua} rule out $m_{\chi} \lesssim \mathcal{O}(10\,{\rm GeV})$.
Therefore,  
pseudo-Dirac fermions with a small mass splitting, Majorana fermions, and scalars, which all have velocity suppressed annihilation cross sections, are considered in this review.

\subsection{Dark Matter Self-Interactions}

Whether dark matter experiences any forces other than gravity, known as self-interacting dark matter (SIDM), is a hotly debated topic in astrophysics.
This section provides a brief overview; the reader is encouraged to see Ref.~\cite{Tulin:2017ara} for a thorough review.
One motivation for SIDM is that the lightest dark matter particle charged under a dark-sector interaction must be stable due to charge conservation, consistent with the fact that dark matter particles have survived for over 14 billion years to date.
SIDM models also have observable implications for astrophysical structure.
For example, self-interactions could explain several small-scale structure observations that appear to be in tension with collisionless dark matter predictions, {\em e.g.}\ the so-called core-cusp problem; however, at large scales, collisionless dark matter models have been a great success.
This can all be reconciled if dark matter self-interactions are velocity dependent, which is expected if 
they are 
mediated by a relatively light $\mathcal{O}($MeV--GeV$)$ mediator particle, {\em e.g.}\ a dark photon.
Indeed, simple dark-photon models can explain observations spanning a large range of length scales, from dwarf galaxies to galaxy clusters.
That said, the interplay between baryonic interactions and dark matter, {\em e.g.}\ how feedback from supernovae affect the dark matter density profile, is not fully understood, and this non-linear dynamic may provide an alternative solution to small-scale structure problems.
Regardless, SIDM is well motivated and searches for dark photons are of great interest to both the particle physics and astrophysics communities.

%% file: strategy.tex
This section describes broadly the strategies experiments use to search for dark photon decays into visible SM and (nearly) invisible dark-sector final states. Since the signatures for these two decay paths are markedly different, the experimental methods used to produce and detect the dark photon signatures can also be quite different.  Below we will summarize the experimental strategies used to search for dark photons and the advantages and challenges of each strategy.

\subsection{Visible Dark Photon Decays}
\label{sec:visstrat}

In scenarios where \aprime decays  into invisible dark-sector final states are kinematically forbidden, the dark photon will decay to visible SM final states. 
If the kinetic mixing strength is  $\varepsilon \gtrsim \mathcal{O}(10^{-3})\times (10\,{\rm MeV} /m_{\aprime})$, the dark photon will decay promptly; {\em i.e.}\ at a point that is experimentally indistinguishable from the production point.  
For smaller mixing strengths, the dark photon decay point may be significantly displaced from the production point.  
Both of these signatures are discussed below.

\subsubsection{Prompt Dark Photon Decays}
\label{sec:prompt}

Searches for prompt visible \aprime decays exploit the fact that the natural width of the dark photon is negligible compared to the mass resolution of any experiment. 
Therefore, such decays will produce a peaking resonant structure, or bump, in the invariant mass spectrum of the decay products, whereas the backgrounds to these searches predominantly produce spectra without sharp features.   
Experiments typically focus on the $e^+e^-$ or $\mu^+\mu^-$ decay channels, since these final states have substantial branching fractions and are the easiest to identify and trigger on.
The backgrounds for these final states largely arise from $\gamma^*\to \ell^+\ell^-$ processes, which are experimentally indistinguishable from the signal making them irreducible. 
In addition, since dark photons couple to the EM current, the expected \aprime yield is related to the observed $\gamma^*\to \ell^+\ell^-$ yield in a small $\pm\Delta m$ window around $m_{\aprime}$ by $n(\aprime \to \ell^+\ell^-) = \varepsilon^2 n(\gamma^* \to \ell^+\ell^-) \mathcal{F}(m_{\aprime})/2\Delta m$, where $\mathcal{F}$ is a known mass-dependent function~\cite{Ilten:2016tkc}.  
Most experiments also have other background sources, including  misidentified particles and leptons produced in weak decays, that also decrease the sensitivity of the analysis.
Many past, existing, and future experiments either have or plan to search for prompt dark photon decays.
The keys to improving sensitivity include increasing the luminosity and improving the mass resolution, which would decrease the effective background yields, both of which are extremely challenging to substantially improve upon. 

\subsubsection{Displaced Dark Photon Decays}
\label{sec:displaced}

Searches for dark photon decay points that are significantly displaced from the production point are able to highly suppress, or even eliminate, the prompt $\gamma^* \to \ell^+\ell^-$ backgrounds.  
Since the dark photon lifetime scales as $[\varepsilon^2 m_{\aprime}]^{-1}$, it is only possible to exploit this displaced-decay signature for small masses and kinetic-mixing strengths.
In addition, low-mass dark photons can be highly boosted resulting in \aprime flight distances up to $\mathcal{O}(100\,{\rm m})$ for sufficiently small values of $\varepsilon$.
In such cases, backgrounds can be highly suppressed by inserting shielding material between the production point and instrumented decay region, a strategy employed by so-called beam-dump experiments. 
For intermediate \aprime lifetimes, where  $c\tau_{\aprime}$ is $\mathcal{O}({\rm mm})$ or less, resolving the \aprime decay point as displaced requires installing a high-precision vertex detector as close as possible to the production point.  
The boost imparted to the \aprime is also important to ensuring that its decay is significantly displaced from the production point.
The primary background in these searches for intermediate-lifetime dark photons are SM photons that convert to $\ell^+\ell^-$ in the vertex detector itself, though via precision {\em in situ} spatial mapping of the detector material, these backgrounds can be highly suppressed by rejecting dilepton vertices that are consistent with locations occupied by material (see, {\em e.g.}, Ref.~\cite{LHCb-DP-2018-002}).
This intermediate-lifetime area is the focus of much ongoing experimental effort.
The keys to improving sensitivity again include increasing luminosities, along with either decreasing the length of the required shielding in beam-dump experiments or improving the vertex resolution at other experiments. 
These are challenging given the suite of high-quality experiments already performed.

\subsection{Invisible Dark Photon Decays}
\label{sec:invisstrat}

If $m_{\aprime} > 2 m_{\chi}$, where $\chi$ again denotes the lightest dark-charged particle, the dark photon will predominantly undergo the (nearly) invisible $\aprime \to \chi\bar{\chi}$ decay. 
We can search for these types of decays in two ways: 
by looking for an excess of events with a consistent invariant mass formed from the imbalance of observed energy and momentum, the {\em missing} mass; 
or 
by looking for the incredibly rare interactions of the dark-sector $\chi$ particles in a detector downstream of the \aprime decay point, referred to as direct detection.  
We will address both of these classes below.

\subsubsection{Missing Mass}
\label{sec:missing}

Experiments that employ the missing-mass method attempt to detect and measure all visible final state particles in each individual interaction, searching for events with an imbalance of energy and momentum. It is important to avoid final states with substantial SM backgrounds from processes that produce neutrinos, as these can mimic a signal.   
Clearly these experiments must also have a precisely determined initial state, which is achieved either  using information from the accelerator or by measuring the energy and momentum of some produced initial state directly. 
For example, the initial state could consist of two beam particles at a collider, or one beam particle striking a stationary, or {\em fixed}, target particle.
Eliminating losses of particles or energy due to gaps in the detector acceptance or the lack of containment of secondary particles is the key to reducing backgrounds, and hence, achieving good sensitivity in these experiments.
Therefore, these detectors  must have extensive and hermetic tracking and EM and hadronic calorimetric systems in order to minimize losses.  
Assuming these losses are minimized, the predominant backgrounds arise from photo-nuclear effects in the calorimeter systems.
Increasing the luminosity while maintaining excellent background suppression is extremely challenging. 

\subsubsection{Direct Detection}
\label{sec:direct}

It is also possible to search for the rare interactions of the dark-sector $\chi$ particles in a detector placed downstream of the \aprime decay point (making the term {\em invisible} somewhat of a misnomer).  
For these experiments, the dark photon is produced in the usual ways, and then subsequently produces the $\chi$ particles in $\aprime \to \chi\bar{\chi}$ decays. 
These $\chi$ particles are then detected via their scattering off of the electrons and nuclei in a downstream detector.  
This scattering involves the $\chi$ particle emitting a dark photon which interacts with the detector particles via its coupling to the EM current, leading to a detectable energy transfer.  
This emission of a dark photon and its coupling to the detector particle leads to an addition suppression factor of $\alpha_D \varepsilon^2$, such that rates in direct-detection experiments scale as $\alpha_D \varepsilon^4$ (compared to $\varepsilon^2$ for other types of experiments).

Due to the substantial suppression factor, direct-detection experiments must be capable of producing a huge number of dark photons, and employ a large active detector mass while maintaining  small background rates.  
Beam-dump experiments are ideal for the upstream component, given their high intensities and low backgrounds.
The large-mass downstream detectors must have sensitivity to low energy transfers, since the recoils of their electrons (protons) are at most $\mathcal{O}(100 \, (10) \, {\rm MeV})$ using existing and near-future beams.
Short-baseline neutrino experiments (or near detectors at long baselines) are well suited to performing these searches; however, one obvious drawback is that they are designed to maximize the neutrino rate in the downstream detector, which sources a large background for $\chi$ searches.  
To overcome this, the MiniBooNE experiment ran with a dedicated direct-detection configuration and achieved good results, see Sec.~\ref{sec:invisible}. 
Another drawback of neutrino experiments is that they typically use proton beams, which by design, produce substantial neutrino backgrounds.
Future dedicated direct-detection experiments propose using electron beams instead, which would greatly reduce the backgrounds.

%% file: visible.tex
\begin{figure}[t]
  \centering
  \includegraphics[width=0.99\textwidth]{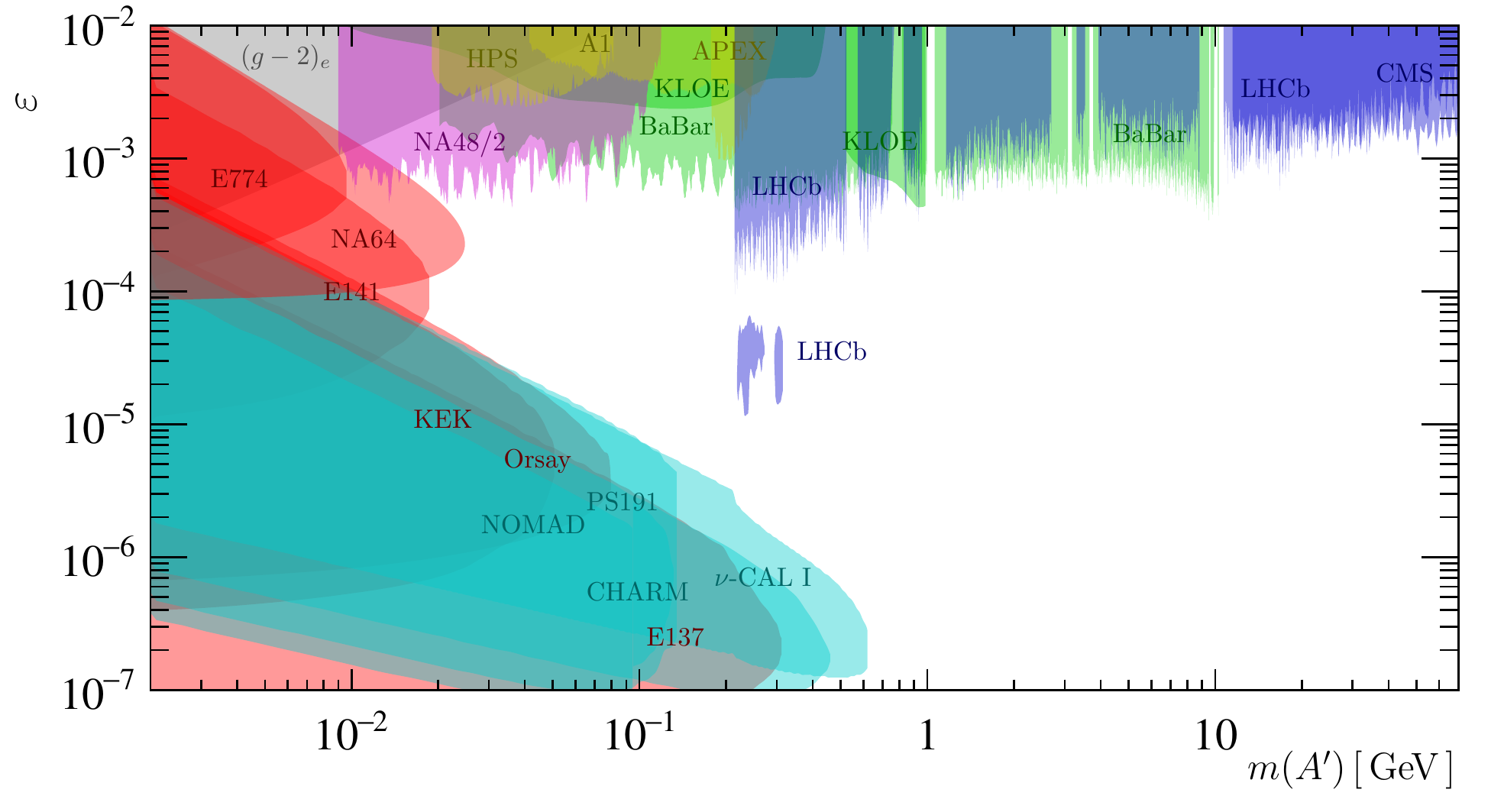}
  \caption{
	From Ref.~\cite{Ilten:2018crw} made using Ref.~\cite{darkcast}:
  Constraints on visible \aprime decays from
  \textcolor{ebdcolor}{electron beam dumps},
  \textcolor{pbdcolor}{proton beam dumps},
  \textcolor{eecolor}{$e^+e^-$ colliders},
  \textcolor{ppcolor}{$pp$ collisions},
  \textcolor{mdcolor}{meson decays},
  and  
  \textcolor{eftcolor}{electron on fixed target} experiments.
  The constraint derived from $(g-2)_e$ is shown in grey~\cite{Pospelov:2008zw,Endo:2012hp}. The gaps in the prompt limits correspond to regions near the masses of the QCD vector mesons.
  }
  \label{fig:visible-limits}
\end{figure}
\begin{figure}[t]
  \centering
  \includegraphics[width=0.99\textwidth]{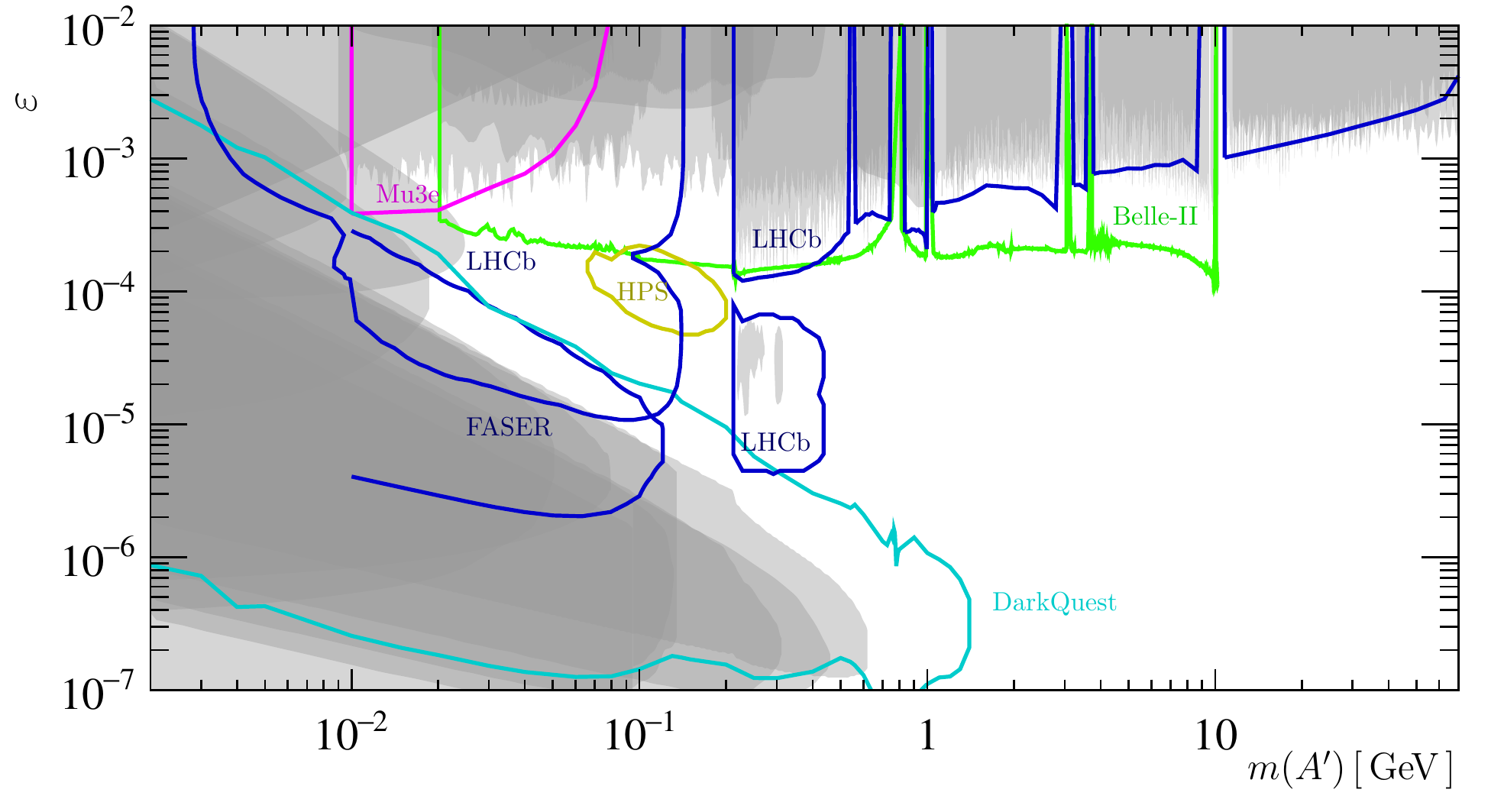}
  \caption{
  Same as Fig.~\ref{fig:visible-limits} but with all existing constraints shown in grey, and proposed future sensitivities in the next 5 years shown as lines (using the same experiment-type color scheme as Fig.~\ref{fig:visible-limits}).
  Only published projections are shown, which for example, does not include a Run~3 update from CMS or the sensitivity of an inclusive $\aprime \to e^+e^-$ search at LHCb.
  }
  \label{fig:visible-limits-future}
\end{figure}

The current constraints on visible \aprime decays in the $[m_{\aprime},\varepsilon]$ plane are presented in Fig.~\ref{fig:visible-limits}. 
The few-loop $\varepsilon$ region is excluded in the low-mass $m_{\aprime} \lesssim 20$\,MeV region. 
For intermediate masses, $0.02 \lesssim m_{\aprime} \lesssim 0.5$\,GeV, there is a gap in the current coverage of roughly $10^{-5} \lesssim \varepsilon \lesssim 10^{-3}$. 
Above 0.5\,GeV, existing results only require $\varepsilon \lesssim 10^{-3}$. 
Projections of the sensitivity expected in the next 5 years from many experiments are shown in Fig.~\ref{fig:visible-limits-future}.
Assuming these are realized, the entire intermediate-mass few-loop region could be explored in the near future; however, there are currently no known viable ways to explore the $m_{\aprime} \gtrsim 1$\,GeV and $\varepsilon \lesssim 10^{-4}$ region. 
This section summarizes the landscape of searches for visible \aprime decays. More details on each experiment are provided in Appendix~\ref{sec:experiments}.

\subsection{Searches in $e^+e^-$ Colliders}

At  $e^+e^-$ colliders, the predominant production mode for dark photons is $e^+e^- \to \aprime \gamma$, where one photon in the SM annihilation process is replaced with a dark photon. 
Thus far, searches at $e^+e^-$ colliders \cite{Lees:2014xha,Ablikim:2017aab,Anastasi:2015qla,Anastasi:2018azp} have looked for prompt \aprime decays in the $e^+e^-$, $\mu^+\mu^-$, and $\pi^+\pi^-$ final states.
The background for these searches is dominated by the irreducible $e^+e^- \to \ell^+\ell^-\gamma$ continuum.  
In addition, there are sizable peaks in the background spectrum due to vector meson decays ($\omega$, $\phi$, J/$\psi$, $\psi(2S)$, $\Upsilon$, {\em etc.}); 
these peak regions are removed from consideration during the searches.
For the $\aprime \to e^+e^-$ decay, there is also background from photon conversions in the detector material, which can be highly suppressed by rejecting $e^+e^-$ vertices that are inconsistent with originating from the primary beam-beam interaction point; however, at low $e^+e^-$ invariant masses, where the opening angle between the leptons is small, some of this conversion background remains. 

As shown in Fig.~\ref{fig:visible-limits}, the BaBar experiment at SLAC has produced the most stringent limits on visibly decaying dark photons in the 1--10\,GeV mass region, and also for 0.1--0.2\,GeV masses in the prompt regime~\cite{Lees:2014xha}.   
BaBar ran from 1999--2008 at center-of-mass collision energies in the range 10.3--10.5\,GeV. 
In addition, Fig.~\ref{fig:visible-limits-future} shows that the Belle~II experiment, which recently started taking data and runs at a similar center-of-mass energy as BaBar, is expected to be sensitive to $\varepsilon$ values that are about 3 times smaller than BaBar~\cite{Kou:2018nap}.  
Looking further down the road, a Future Circular Collider could produce world-leading sensitivity up to higher masses~\cite{Karliner:2015tga,He:2017zzr,DOnofrio:2019dcp}.  
These results and projections demonstrate that $e^+e^-$ colliders are powerful probes of prompt visible dark photon decays. 

\subsection{Searches in Hadron Colliders}

At  the LHC, the predominant production modes for dark photons depend on $m_{\aprime}$: 
meson decays, {\em e.g.}, $\pi^0 \to \aprime \gamma$ and $\eta \to \aprime \gamma$, 
for $m_{\aprime} \lesssim 0.5$\,GeV;
\aprime mixing with the $\rho$, $\omega$, and $\phi$ mesons for $0.5 \lesssim m_{\aprime} \lesssim 1$\,GeV;
and Drell-Yan, $q\bar{q} \to \aprime$, for larger masses. 
Thus far, searches at the LHC have looked for both prompt (LHCb and CMS) and displaced (LHCb) $\aprime \to \mu^+\mu^-$ decays, with plans to also search for $\aprime \to e^+e^-$ decays at LHCb.
The background for the prompt searches is dominated by the irreducible SM Drell-Yan continuum, leptons produced in heavy-flavor decays, and hadrons misidentified as leptons.  
The sizable peaks  
due to QCD vector meson decays are removed from consideration during the searches.
The background in the search for displaced \aprime decays is dominated by: photon conversions, which can be highly suppressed by rejecting dilepton vertices that are consistent with originating from locations occupied by detector material; leptons produced in heavy-flavor decays, though these are only important for \aprime lifetimes $\mathcal{O}({\rm ps})$; and $K_S \to \pi^+\pi^-$ decays, where both pions are misidentified as leptons, for $0.35 \lesssim m_{\aprime} \lesssim 0.5$\,GeV. 

Figure~\ref{fig:visible-limits} shows that the LHCb~\cite{LHCb-PAPER-2017-038,Aaij:2019bvg} and CMS~\cite{Sirunyan:2019wqq} experiments have produced the most stringent limits on visibly decaying dark photons for $m_{\aprime} > 10$\,GeV. 
In addition, LHCb has produced world-leading constraints for $2m_{\mu} \lesssim m_{\aprime} \lesssim 0.5$\,GeV from both its prompt and displaced searches, the latter being the only search to achieve sensitivity %
using a displaced-vertex signature. 
This is made possible by its high-precision vertex detector and flexible trigger system. 
Figure~\ref{fig:visible-limits-future} shows that LHCb is expected to greatly enhance its sensitivity in the next 5 years, specifically with its upgrade for LHC Run~3~\cite{Ilten:2015hya,Ilten:2016tkc}. 
LHCb also plans to perform inclusive searches for $\aprime \to e^+e^-$ decays, using both current and future data, which should explore the mass gap between its published projections, 
and CMS expects to substantially improve on its sensitivity in the high-mass region as well, though no published projections exist for either of these searches. 
These results demonstrate that LHC experiments are powerful probes of both prompt and displaced visible \aprime decays. 

Several new experiments have been proposed recently that will search for dark photons produced in LHC proton-proton collisions using new detector packages dedicated to long-lived \aprime scenarios. 
For example, FASER~\cite{Feng:2017uoz} is being installed in an LHC service tunnel positioned 480\,m downstream from the ATLAS interaction point. 
FASER plans to start collecting data in LHC Run~3 and expects to provide the \aprime sensitivity~\cite{Ariga:2018uku}  shown in Fig.~\ref{fig:visible-limits-future}.
An upgrade, FASER2, has been proposed, that would increase the luminosity by a factor of 20, greatly increasing the \aprime sensitivity. 
Similar experiments like MATHUSLA~\cite{Curtin:2018mvb} and Codex-b~\cite{Gligorov:2017nwh} could also come online in the next decade. 

\subsection{Searches in Electron Beam Fixed Target Experiments}

Fixed-target experiments send a high-current electron beam through a thin, high-$Z$ target (or assembly of targets) and detect the products with a downstream, forward detector. Production of dark photons is via a radiative \aprime recoiling against the target nucleus, {\em i.e.}\ via the $eZ \to eZ\aprime$ bremsstrahlung process~\cite{Bjorken:2009mm,Beranek:2013yqa}.  
These radiative dark photons are largely produced in the forward region with close to the beam energy, leading to decay leptons that are boosted in the forward direction and carry (on average) half of the beam energy.  There is a large background from QED trident events, of which the radiative part is irreducible.  
As discussed in Sec.~\ref{sec:prompt}, since \aprime production is proportional to the SM $\gamma^* \to e^+e^-$ yield, these experiments are normalized using the observed number of radiative events. Therefore, the total integrated luminosity, which typically has large uncertainties in this type of experiment, does not need to be determined.  

The A1 (Mainz Microtron)~\cite{Merkel:2014avp} and APEX (JLAB)~\cite{Essig:2010xa,Abrahamyan:2011gv} experiments both utilized movable, low-angle, high-resolution spectrometers to search for prompt $\aprime\to e^+e^-$ decays.   
The HPS experiment (JLAB) instead uses a silicon vertex tracker inside of a magnetic field near the target 
which enables searching for both prompt and displaced \aprime decays~\cite{Moreno:2013mja,Adrian:2018scb}. 
Figure~\ref{fig:visible-limits} shows that experiments in this category do not currently provide world-leading sensitivity to dark photons; however, both APEX and HPS have further running planned in the near future, and expect to expand their reach into unexplored regions of \aprime parameter space (see Fig.~\ref{fig:visible-limits-future}).  
Similar experiments have been proposed for running further into the future~\cite{Freytsis:2009bh,
Balewski:2013oza,Raggi:2014zpa,Nardi:2018cxi}.

\subsection{Searches in Electron Beam Dumps}

Data from the electron beam dump experiments E141, E137, E774, KEK, and Orsay~\cite{Riordan:1987aw,Bjorken:1988as,Bross:1989mp,Konaka:1986cb,Davier:1989wz}, which ran during the 1980s, have been recast~\cite{Bjorken:2009mm,Andreas:2012mt} to place constraints on displaced visible $\aprime \to e^+e^-$ decays. 
These experiments exploit the same $eZ \to eZ\aprime$ bremsstrahlung production process as the experiments in the previous subsection.
A major difference, however, is that they all employed shielding after their beam-dump targets to absorb SM particles. 
Each experiment employed an EM calorimeter placed downstream of its substantial shielding.
The length of the shielding sets a hard lower limit on the \aprime flight distance in the lab frame. 
Since the \aprime lifetime scales as $[\varepsilon^2 m_{\aprime}]^{-1}$, beam-dump experiments provide sensitivity to wedge-shape regions of dark-photon parameter space. 
There is also one experiment in this category that is currently running: NA64 uses a 100\,GeV electron beam derived from the CERN SPS proton beam incident on an active target upstream of an EM calorimeter to enable searches for dark photon decays to $e^+e^-$~\cite{Banerjee:2019hmi}.   
Figure~\ref{fig:visible-limits} shows that electron beam dump experiments provide world-leading constraints on dark photons in the low-mass region. 
Future experiments have also been proposed; see, {\em e.g.}, Ref.~\cite{Seo:2020dtx}.

\subsection{Searches in Proton Beam Dumps}

Data from previously run proton beam dump experiments have also been used to place limits on $\aprime \to e^+e^-$ decays.
The experimental setups were similar to those in the previous subsection; however, the use of a proton beam results in more possible production processes to consider. 
Thus far, limits have been set by the following experiments: $\nu$-CAL~I~\cite{Blumlein:1990ay,Blumlein:1991xh}, using $\pi^0\to\aprime\gamma$ decays~\cite{Blumlein:2011mv} and proton bremsstrahlung~\cite{Blumlein:2013cua}; 
CHARM~\cite{Bergsma:1985qz}, using $\eta^{(\prime)}\to\aprime\gamma$ decays~\cite{Gninenko:2012eq};
and NOMAD~\cite{Astier:2001ck} and PS191~\cite{Bernardi:1985ny} using $\pi^0\to\aprime\gamma$ decays~\cite{Gninenko:2011uv}.
Figure~\ref{fig:visible-limits} shows that these limits are largely comparable to those obtained from electron beam dumps, exceeding them at higher masses.
In the near future, 
two variations of the SeaQuest experiment, dubbed SpinQuest and DarkQuest, will use the 120\,GeV main injector proton beam at Fermilab incident on a beam dump to search for $\aprime \to \mu^+\mu^-$ and $\aprime \to e^+e^-$ decays~\cite{Gardner:2015wea,Tsai:2019mtm}. 
As can be seen in Fig.~\ref{fig:visible-limits-future}, this will greatly enhance the sensitivity compared to that of the older proton beam dumps. 
Further into the future, proposed experiments like SHiP could provide even greater enhancements~\cite{Alekhin:2015byh}. 

\subsection{Searches in Meson and Lepton Decays}

While meson-decay processes contribute to the production of dark photons at hadron colliders and proton beam dumps, experiments like NA48/2, located at the CERN SPS, exclusively exploit such decays. 
Specifically, NA48/2 searched for $\pi^0 \to \aprime \gamma$ followed by prompt $\aprime \to e^+e^-$ decays using $\pi^0$ mesons produced in $K^+ \to \pi^+ \pi^0$ decays~\cite{Batley:2015lha}.  
Figure~\ref{fig:visible-limits} shows that the NA48/2 constraints are world leading for prompt decays in the 10--100\,MeV mass region.
Soon, the Mu3e experiment located at the Paul Scherrer Institute expects to provide the first dark-photon sensitivity in lepton decays using stopped muons~\cite{Echenard:2014lma}. 
Figure~\ref{fig:visible-limits-future} shows that Mu3e could be sensitive to currently unexplored parameter space soon. 
Future experiments that exploit other meson decays are also being considered; see, {\em e.g.}, Ref.~\cite{Gan:2020aco}.

%% file: invisible.tex
\begin{figure}[t]
  \centering
  \includegraphics[width=0.99\textwidth]{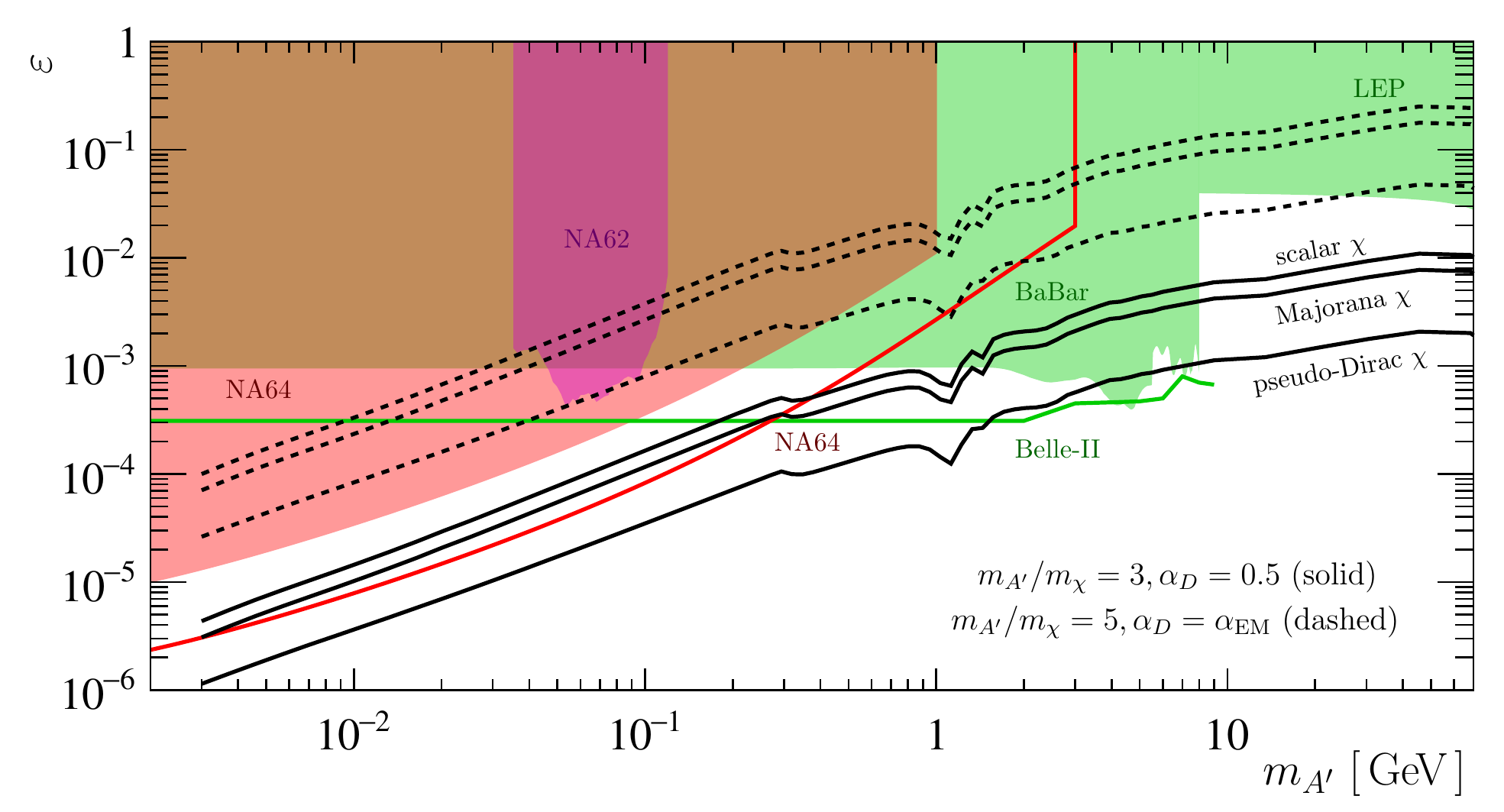}
  \caption{
  Updated from Ref.~\cite{Ilten:2018crw} using Ref.~\cite{darkcast}:
  Constraints and proposed sensitivity within the next 5 years on invisible \aprime decays. The color scheme is the same as in Fig.~\ref{fig:visible-limits}. The (solid and dashed black lines) thermal targets for different dark matter scenarios are also shown~\cite{Akesson:2018vlm}. 
  }
  \label{fig:invisible-limits}
\end{figure}

The current constraints on invisible $\aprime \to \chi\bar{\chi}$ decays are summarized in Fig.~\ref{fig:invisible-limits}.
These results were obtained by looking for an excess of events with a consistent missing invariant mass, formed from the imbalance of observed energy and momentum.
In this section, we will first discuss how these results were obtained, and the near-term prospects for improvement. 
Then, in Sec.~\ref{sec:invDM} we will discuss the alternative strategy of direct detection of the incredibly rare interactions of the dark-sector $\chi$ particles in a detector downstream of the \aprime decay point.
Figure~\ref{fig:invisible-limits} shows that existing constraints exclude otherwise viable thermal dark matter scenarios, {\em e.g.}\ EM-like values of $\alpha_D$.
Furthermore, even pessimistic scenarios with a large $\alpha_D$ and small $m_{\aprime} / m_{\chi}$ ratio will be accessible in the near future. 

\subsection{Searches at $e^+e^-$ Colliders}

As discussed above, the predominant production mode for dark photons in these experiments is $e^+e^- \to \aprime \gamma$, where one photon in the SM annihilation process is replaced with a dark photon. 
For the case where the \aprime decays invisibly, the visible final state is a single photon with a substantial imbalance in momentum and energy compared to the initial $e^+e^-$ beams. 
The dominant backgrounds are $e^+e^- \to \gamma \gamma$, where one photon is not detected, $e^+e^- \to \gamma \gamma \gamma$, with one photon out of acceptance and another undetected, and  $e^+e^- \to e^+e^-\gamma$, radiative Bhabha scattering where both leptons are out of acceptance.  
The $e^+e^- \to \gamma \gamma$ background mimics a low-mass \aprime signal. 
In the BaBar experiment, the primary cause of not detecting a photon was azimuthal gaps in the calorimeter crystals~\cite{Lees:2017lec}.  
This was highly suppressed by looking for subsequent interactions of the photon in the muon system, and by disfavoring missing momenta that are consistent with the locations of these calorimeter gaps.  
The  $e^+e^- \to \gamma \gamma \gamma$ background is dominant at intermediate masses, while 
the radiative Bhabha background is irreducible and dominant for $m_{\aprime} \gtrsim 4$\,GeV. 

Figure~\ref{fig:invisible-limits} shows that the BaBar constraints on invisible \aprime decays are world leading in the 0.2--8\,GeV mass region~\cite{Lees:2017lec}.   
The structure in the BaBar constraints at higher masses is due to the fact that its singe-photon-trigger data was collected at several center-of-mass energies, corresponding to the $\Upsilon(2S)$, $\Upsilon(3S)$, and $\Upsilon(4S)$ masses, and with two different energy thresholds for the detected SM photon. 
Belle~II projections~\cite{10.1093/ptep/ptz106} indicate that competitive sensitivity with BaBar can be achieved with almost 5 times less data.  
There are two aspects of the Belle~II calorimeter design that enhance its sensitivity to invisible dark photons: 
larger solid angle coverage due to both a larger calorimeter and a smaller beam-energy asymmetry; 
and the crystals in the central (barrel) part of the calorimeter are tilted so that gaps between crystals do not align with the interaction point. 
We note that the projections in Fig.~\ref{fig:invisible-limits} are for only $\approx 0.04\%$ of the full planned Belle~II data sample; extrapolations to the full sample require studies of the systematic uncertainties in photon detection probabilities. 
The KLOE2 experiment has also recorded a 
data set with a single photon trigger at a much smaller center-of-mass energy than BaBar and Belle~II,  
though sensitivity projections are not yet available.
Finally, for masses larger than 10\,GeV, 
mono-photon data from the DELPHI experiment at LEP~\cite{Abdallah:2003np,Abdallah:2008aa}, which was recast as a search for invisible dark photon decays in Ref.~\cite{Fox:2011fx}, provides the strongest constraints.

\subsection{Searches at Electron Beam Dumps}

The aforementioned NA64 experiment, discussed in Sec.~\ref{sec:visible}, is also able to search for invisible \aprime decays. 
Dark photons would be produced via hard bremsstrahlung from its 100\,GeV electron beam interacting in an active target, an EM calorimeter. 
The subsequent $\aprime \to \chi\bar{\chi}$  decay would carry off substantial energy, resulting in an energy deposition in the  EM calorimeter well below the beam energy. 
The vast majority of SM events with small EM energy deposited in the calorimeter involve the production of hadrons, which deposit substantial energy in the hadronic calorimeter located further downstream. 
After rejecting events with activity in the hadronic calorimeter, the largest surviving background is due to electro-production of a hadron in the beam line prior to the beam electron reaching the EM calorimeter, where the hadron misses the hadronic calorimeter.  %
These NA64 limits are the most stringent available for $m_{A^\prime}<0.2$~GeV~\cite{NA64:2019imj}. 
If the analysis remains background free, the limits on $\varepsilon$ will improve as the square root of the number of electrons on target. 
Projections for the full data sample are also shown in Fig.~\ref{fig:invisible-limits}, while further improvements from running with a muon beam are also possible~\cite{Gninenko:2019qiv}. 

LDMX is a proposed electron beam fixed target experiment that would search for the invisible decay of dark photons~\cite{Akesson:2018vlm}. In addition to EM and hadronic calorimetry, the detector includes tracking in a magnetic field before and after the thin target, enabling measurements of both missing momentum and missing energy. This additional information provides better rejection of rare backgrounds than energy measurements alone.  LDMX expects to provide good coverage of the most pessimistic relic density targets for dark photon masses up to hundreds of MeV. Various sites are under consideration, including a proposed new beam line derived from the LCLS beam at SLAC.

BDX is a proposed experiment to be located in a new underground experimental hall downstream of the JLAB Hall A beam dump~\cite{Battaglieri:2016ggd}. It will use a homogeneous CsI(Tl) calorimeter to detect $\chi$ particles scattering off of electrons. %
BDX expects to probe unexplored parameter space for dark photon masses up to several hundred MeV. 

\subsection{Searches in Meson Decays}

The NA62 experiment has exploited its large sample of charged kaons and its hermetic photon coverage to search for invisible dark photons produced in $\pi^0 \to \gamma A^\prime$ decays. 
The $\pi^0$ mesons are produced in $K^+ \to \pi^+ \pi^0$ decays, where 
the four-momenta derived from that of the $K^+$ and $\pi^+$ must be consistent with a $\pi^0$.
Dark photon candidates are the subset of these events that contain only a single detected photon.  
To suppress backgrounds from photon conversions, the only hit in a 
scintillator upstream of the EM calorimeter must be from the $\pi^+$.
The dominant remaining background is $K^+ \to \pi^+ \pi^0 (\gamma)$ where one of the $\pi^0 \to \gamma\gamma$ photons is lost due to a photonuclear reaction or conversion.
The resulting limits~\cite{CortinaGil:2019nuo}, which are shown in  Fig.~\ref{fig:invisible-limits}, are comparable to NA64, but not world leading; however, we note that these results used only about 1\% of the full NA62 data sample. 
Finally, as noted in Ref.~\cite{Beacham:2019nyx}, the proposed KLEVER experiment~\cite{Ambrosino:2019qvz} may be able to probe invisible decays of dark photons with masses in the 100--200~MeV range in conjunction with its study of the rare decay $K_L\to \pi^0 \nu \bar \nu$.

\subsection{Searches for Dark Matter Produced in Proton Beam Dumps}
\label{sec:invDM}

\begin{figure}[t]
  \centering
  \includegraphics[width=0.99\textwidth]{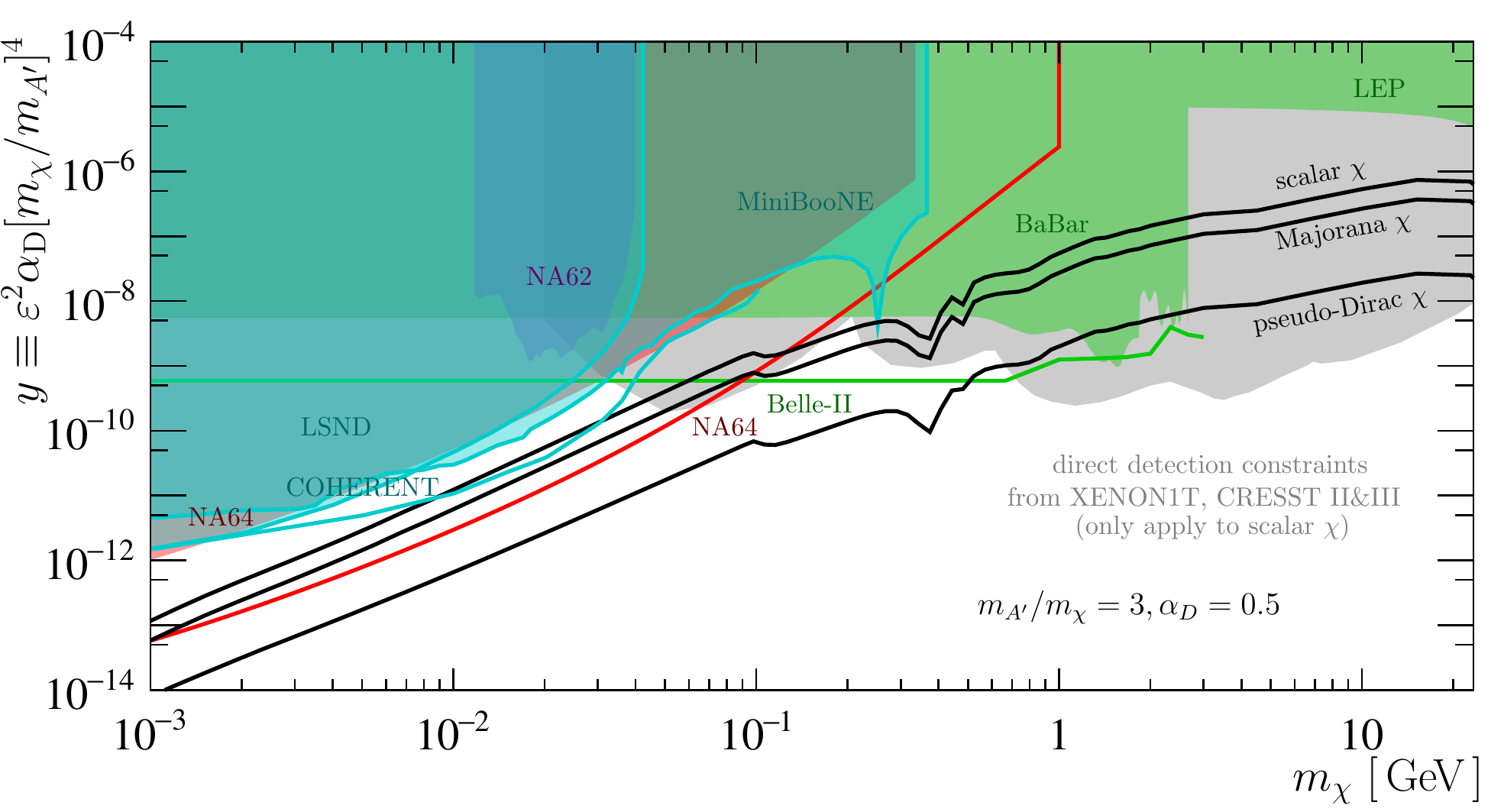}
  \caption{
  Constraints on the invisible $\aprime \to \chi \bar{\chi}$ scenario in the $[m_{\chi},y]$ plane for pessimistic values of $\alpha_D$ and the $m_{\aprime} / m_{\chi}$ ratio.
  All of the results from Fig.~\ref{fig:invisible-limits} also apply here. 
  In addition, searches for dark photons produced in proton beam dumps with the subsequent scattering of $\chi$ particles detected downstream are shown using the same \textcolor{pbdcolor}{proton beam dump} color scheme as before. The direct detection constraints from CRESST~II~\cite{Angloher:2015ewa}, CRESST~III~\cite{Abdelhameed:2019hmk}, and XENON1T~\cite{Aprile:2019xxb} only apply for the elastic scalar dark matter case and have sizable uncertainties~\cite{Baxter:2019pnz,Kurinsky:2020dpb}. Otherwise, direct detection experiments have much worse sensitivity than accelerator searches, since $\chi$ particles produced at accelerators are relativistic. As in Fig.~\ref{fig:invisible-limits}, less pessimistic values of dark-sector parameters are already excluded.  
  }
  \label{fig:invisible-limits-DM}
\end{figure}

An alternative strategy for detecting $\aprime \to \chi \bar{\chi}$ decays is to detect the incredibly rare interactions of the dark-sector $\chi$ particles in a detector downstream of the \aprime decay point.
Given how unlikely such interactions are, direct-detection experiments must be capable of producing a huge number of dark photons, and employ a large active detector mass while maintaining small background rates.  
Proton beam dumps placed upstream of neutrino detectors are well suited to performing these searches. 
Dark photons are predominantly produced at proton beam dumps via $\pi^0$ and $\eta$ decays, along with proton bremsstrahlung. 
The $\chi$ particles can be detected through neutral-current-like processes, $e\chi \to e\chi$, $N\chi \to N\chi$, or if the $\chi$ beam is energetic enough, scattering off of nucleons can also create pions. 
Charged-current quasi-elastic neutrino scattering, which has no analog involving dark matter interactions, can be used to normalize the neutrino flux. 

The LSND experiment produced a huge sample of $\pi^0$ decays by impinging a low-energy proton beam onto a fixed target.
Any subsequent $\aprime \to \chi \bar{\chi}$ decays would produce low-energy $\chi$ particles only capable of transferring a visible amount of energy to electrons, thus only $e\chi \to e\chi$ scattering was considered in the reinterpretation of its results~\cite{deNiverville:2011it}.  
Figure~\ref{fig:invisible-limits-DM} shows that the LSND limits are competitive with those obtained from searches for invisible \aprime decays. 

The MiniBooNE experiment could produce $\chi$ particles with sufficient energy to produce visible scatters off of both electrons and nucleons; however, 
the electron-scattering analysis still provides better sensitivity in the minimal model~\cite{Aguilar-Arevalo:2018wea} (the nucleon-scattering analysis is more sensitive to leptophobic models, where the  $e\chi$ coupling is highly suppressed).  
The dominant background is due to neutral-current neutrino scattering events. 
Figure~\ref{fig:invisible-limits-DM} shows that the MiniBooNE limits are also competitive with those obtained from searches for invisible \aprime decays. 

The COHERENT experiment plans to search for dark matter scattering at low momentum transfer, $Q^2 < (50 \, \mathrm{MeV})^2$, where the $\chi$ particles can interact coherently with an entire nucleus in the detector, producing a several keV nuclear recoil~\cite{Akimov:2019xdj}. 
Neutrinos, which are produced by charged pion decays at rest, will produce a similar signal by design. A major difference is that the dark matter signal is prompt (coincident with the 600\,ns long beam bunch), whereas the neutrino signal has both a prompt component and a component delayed by the muon lifetime. The recoil energy spectra are also somewhat different. 
As can be seen in Fig.~\ref{fig:invisible-limits-DM}, COHERENT expects to probe some unexplored parameter space soon.

Figure~\ref{fig:invisible-limits-DM} shows that the thermal targets for accelerator-based experiments have minimal dependence on the nature of the dark matter particles. 
This is because dark matter produced at accelerators is relativistic, and due to the fact that the strength of the \aprime interactions with SM particles are fixed at thermal freeze-out.
Conversely, the rate of non-relativistic relic dark-matter scattering in direct-detection experiments varies by $\approx 20$ orders of magnitude between the elastic scalar and pseudo-Dirac scenarios considered here. 
Therefore, as can be seen in Fig.~\ref{fig:invisible-limits-DM}, accelerator-based experiments can probe nearly all thermal scenarios in this mass regime in the next 5 years, with future efforts fully covering these targets~\cite{Battaglieri:2016ggd,SHiP:2020noy,Doria:2019sux,Akesson:2018vlm}. 
Near-future direct-detection experiments will only be able to explore the elastic scalar case.
That said, while discovery of \aprime and/or $\chi$ particles at an accelerator would undoubtedly be Nobel worthy, these may not be part of the sector that constitutes the majority of the dark matter in the universe. 
Further study would be needed considering astrophysical, cosmological, direct and indirect detection data sets, which demonstrates the complementarity inherent in understanding the nature of dark matter.

%% file: rich.tex
The minimal dark photon model described in Sec.~\ref{sec:theory} is not the only dark-sector option.
The strongest connection to the dark sector may not arise via kinetic mixing.
Furthermore, the dark sector could be populated by other types of particles that have phenomenological implications.
Indeed, studying non-minimal dark sectors is a major challenge due to the wide array of viable dark-sector models.
This section briefly discusses a few examples of {\em rich} dark sectors.

\subsection{Feeble Direct Couplings}

\begin{figure}[t]
  \centering
  \includegraphics[width=0.99\textwidth]{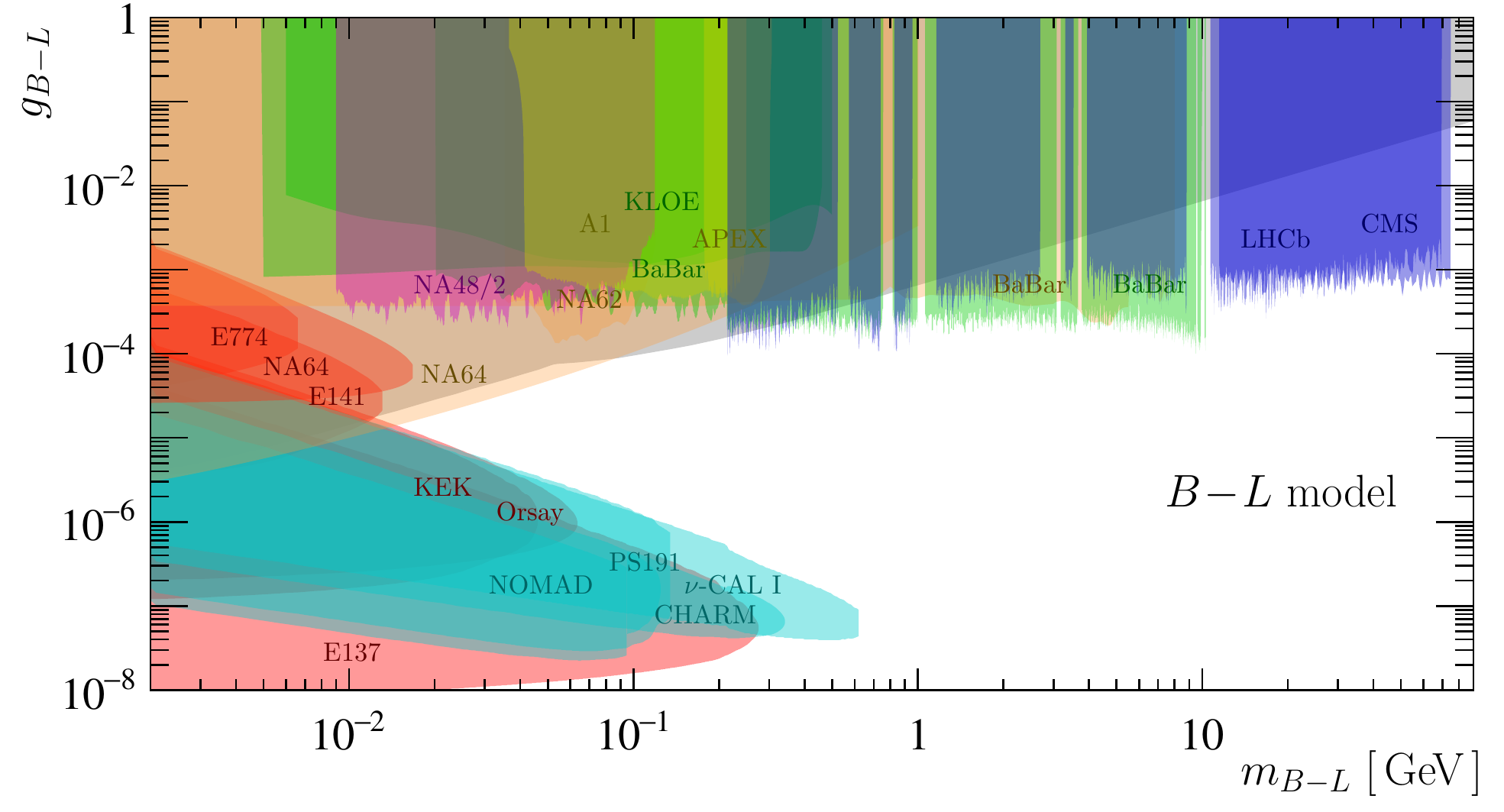}
  \caption{
  Updated from Ref.~\cite{Ilten:2018crw} using Ref.~\cite{darkcast}:
  Constraints derived on \BL decays to SM final states using the same experimental color scheme as in Fig.~\ref{fig:visible-limits}.
  The (orange) invisible constraints also apply to \BL due to its coupling to neutrinos.
  The grey constraints are from the neutrino-scattering experiments Borexino~\cite{Harnik:2012ni,Bellini:2011rx}, Texono~\cite{Bauer:2018onh,Deniz:2009mu}, CHARM-II~\cite{Bauer:2018onh,Vilain:1993kd}, COHERENT~\cite{Cadeddu:2020nbr}.
  }
  \label{fig:B-L-limits}
\end{figure}

The dark-photon portal described in Sec.~\ref{sec:aprime-portal} is not the only way to connect the light and dark sectors.
The SM fields could be charged directly under the gauge interaction of the dark sector, resulting in very different couplings {\em cf.}\ a kinetically mixed dark photon.
For example, \BL interactions would also result in a coupling of the \aprime to neutrinos.
While this alters the results of Sec.~\ref{sec:theory}, a data-driven approach similar to that described above for the dark photon can be used for \BL and any other vector-boson model to determine the production and decay rates~\cite{Ilten:2018crw}.
Furthermore, dark-photon searches provide serendipitous discovery potential for \BL bosons, as well as many other types of new particles~\cite{Ilten:2018crw,Fayet:1990wx,Fayet:2006sp,Fayet:2007ua}.
Figure~\ref{fig:B-L-limits} shows the constraints placed on the \BL model from both dark-photon searches and from neutrino-scattering measurements.
If the \BL boson mass $m_{B-L}$ is below the weak scale, its associated gauge coupling must be tiny.

The \BL model is popular because, exceptionally, it is an anomaly-free global symmetry (assuming only the existence of three right-handed neutrinos).
This is important because classical symmetries can be broken through quantum-mechanical triangle anomalies, which must be canceled to obtain a valid quantum field theory.
Gauging an anomalous current requires introducing new chiral fermions with electroweak charges to cancel the anomalies.
Such models must then address the fact that these fermions have not been observed, neither directly, {\em e.g.}\ at the LHC, nor indirectly, via the enhanced couplings they produce in the low-energy theory between the \aprime and the $W$ and $Z$ bosons, {\em e.g.}\ which enhance the \aprime production rate in penguin decays.
See Ref.~\cite{Dror:2017ehi} for detailed discussion on this topic.

\subsection{Dark Supersymmetry}

Supersymmetric models can also produce hidden sectors at low energy scales, which is often called dark SUSY (see, {\em e.g.}, Ref.~\cite{ArkaniHamed:2008qn}).
In these models, the coupling of the dark photon to SM particles typically still arises via the same kinetic mixing term as in the minimal dark photon model;  therefore, the \aprime decays and lifetime are still given by Sec.~\ref{sec:aprime-decays}.
However, the fact that the dark sector is connected via SUSY to the SM gives rise to additional production mechanisms, {\em e.g.}, involving decays of the Higgs boson.
A standard benchmark case involves the Higgs decaying into the two lightest non-dark neutralinos, followed by each of these decaying into the dark matter particle (presumably a dark neutralino) and a dark photon.
The dark matter particles are undetected while the two dark photons each decay as shown in Fig.~\ref{fig:aprime_decays}, though in practice only the $\aprime \to \mu^+\mu^-$ decays are used.
The detector signature is, thus, two isolated (possibly displaced) pairs of high-$p_{\rm T}$ muons (the large transverse momentum imparted by the large Higgs mass).
Both ATLAS~\cite{Aad:2014yea,Aad:2015sms,Aad:2019tua} and CMS~\cite{Sirunyan:2018mgs} have searched for these types of processes.
Together, they have ruled out almost all of the $\epsilon \gtrsim 10^{-7}$ parameter space for $2\,m_{\mu} \lesssim \ma < 9$\,GeV, under the assumption that $\mathcal{B}(H \to 2\,\aprime + X) \gtrsim 1\%$.

\subsection{Strongly Interacting Dark Matter}

The low-energy phenomenology of the dark sector could be dominated by a strongly coupled interaction, similar to QCD, along with the force mediated by the dark photon (with all SM particles uncharged under both interactions, and the \aprime coupled to the SM via kinetic mixing).
The dark matter in such a scenario would then be the pseudo-Nambu-Goldstone bosons of a spontaneously broken chiral symmetry, {\em i.e.}\ the dark pions of the strongly coupled interaction, which are stable; see, {\em e.g.}, Ref.~\cite{Hochberg:2014kqa}.
As in QCD, the dark vector mesons will also be important phenomenologically, {\em e.g.}, they will mix with the dark photon analogously to $\rho$--$\gamma$ mixing, coupling the dark photon to the dark pions.
These vector mesons can produce striking signals at accelerators in most of the cosmologically favored parameter space, where they are naturally long-lived resulting in missing energy from the dark pions produced along with displaced vertices of SM particles if they decay within the detector.
Dark pion masses from roughly 10\,MeV to $\mathcal{O}({\rm GeV})$ are cosmologically allowed.
The strongest constraints come from the invisible \aprime searches discussed in Sec.~\ref{sec:invisible}, where the dark vector meson must decay outside of the detector, and from visible \aprime searches at beam-dump experiments, where the dark vector meson must decay after the shielding but otherwise within the detector, and from a recent search performed by LHCb~\cite{Aaij:2020ikh}.  
Future searches at colliders and fixed-target experiments that look for the displaced vertex signature, whose topology is markedly different from that of an \aprime decay due to the missing energy and since the \aprime can be off shell, can greatly improve the discovery potential for this class of dark-sector models~\cite{Berlin:2018tvf}.

\subsection{Inelastic Dark Matter}

Another interesting area under study is dark sectors that contain unstable but long-lived particles.
For example, inelastic dark matter models that contain two nearly mass-degenerate states, $\chi_1$ and $\chi_2$ with $m_2 > m_1$, which are coupled off-diagonally to the dark photon leading to the processes $\chi_1 \chi_2 \to A'^* \to {\rm SM}$ and $\chi_2 \to \chi_1 A'^{(*)} \to \chi_1 + {\rm SM}$, where the \aprime again couples to the SM via kinetic mixing. ({\em N.b.}, such a scenario is natural in the strongly coupled class of dark-matter models discussed in the previous section.)
Since the heavier $\chi_2$ state is unstable, dark matter annihilations essentially stop once the $\chi_2$ population has died off, which avoids the otherwise stringent bounds due to CMB anisotropies (assuming the mass splitting provides sufficient phase space to keep the $\chi_2$ lifetime short enough).
Furthermore, constraints from direct detection experiments are severely weakened, since inelastic scatters are kinematically suppressed by the mass splitting and elastic scatters are loop suppressed.
Therefore, the only way to test this scenario is at accelerator experiments, where $\chi_2 \chi_1$ pair production is mediated by a possibly off-shell dark photon.
The characteristic $\chi_2$ decay length is comparable to that of a typical charged-particle tracking (sub)system, resulting in an experimental signature of a displaced $A'^{(*)} \to {\rm SM}$ vertex and missing energy if $\chi_2$ decays in the detector, and only missing energy otherwise.
The best existing constraints come from the invisible \aprime searches discussed in Sec.~\ref{sec:invisible}, and this scenario will also benefit from improved invisible searches.
In addition, dedicated searches that include the displaced vertex signature are planned that will greatly improve the sensitivity at higher masses; see, {\em e.g.}, Refs.~\cite{Duerr:2019dmv,Berlin:2018jbm}.

%% file: sum.tex
\begin{summary}[SUMMARY POINTS]
\begin{enumerate}
\item Dark matter particles may interact with other dark matter particles via a new force mediated by a dark photon, which would be the dark-sector analog to the photon. 
\item The \aprime can obtain a highly suppressed mixing-induced coupling to the EM current,
providing a portal through which dark photons can interact with ordinary matter.
\item The minimal dark-photon model only has 3 unknown parameters:  the strength of the kinetic mixing, $\varepsilon$; the dark photon mass, \ma; and the decay branching fraction of the dark photon into invisible dark-sector final states, which would be either (nearly) unity or zero (corresponding to whether any invisible dark-sector final states are kinematically allowed or not).
\item Great progress has been made recently in exploring the \aprime parameter space.
However, well-motivated scenarios, including the $0.02 \lesssim m_{\aprime} \lesssim 0.5$\,GeV and $10^{-5} \lesssim \varepsilon \lesssim 10^{-3}$ region in the visible \aprime case, and thermal targets for large $\alpha_D$ in the invisible case, remain unexplored. 
\item One striking advantage of producing dark matter in the lab is that it will be relativistic, which leads to accelerator-based experiments having similar sensitivity to most types of dark matter particles. 
This is in stark contrast to direct-detection experiments, which, {\em e.g.}, will only be able to explore thermal targets for the elastic scalar case in the near future. 
That said, an accelerator-based discovery may not be what constitutes the majority of the dark matter in the universe.
Further study would be needed considering astrophysical, cosmological, direct and indirect detection data sets, which demonstrates the complementarity inherent in understanding the nature of dark matter.
\end{enumerate}
\end{summary}

\begin{issues}[FUTURE ISSUES]
\begin{enumerate}
\item For visible dark photons, the entire few-loop $\varepsilon$ region could be explored in the near future for $m_{\aprime} \lesssim 0.5$\,GeV; however, there are currently no known viable ways to explore the $m_{\aprime} \gtrsim 1$\,GeV and $\varepsilon \lesssim 10^{-4}$ region. Will any future experiments manage to probe this space? 
\item For invisible dark photons, existing constraints already exclude many otherwise viable thermal dark matter scenarios. 
Even pessimistic scenarios will be accessible in the near future. Can well-defined targets be defined for dark matter scenarios that do not involve thermal equilibrium in the early universe? 
\item Dark matter self-interactions could explain several small-scale structure anomalies; however, the interplay between baryonic interactions and dark matter is not fully understood, and this may provide an alternative solution. Can enough progress be made in these calculations to enable defining accelerator-based \aprime targets motivated by small-scale structure? 
\item The minimal \aprime scenario is not the only option. Studying non-minimal dark sectors is a major challenge due to the wide array of viable models.
Dark-photon searches provide serendipitous discovery potential for many other, but not all, types of new particles. How can we maximize our exploration capabilities? Are we missing other well-defined targets similar to thermal dark matter for these models? 
\end{enumerate}
\end{issues}

%% file: experiments.tex
This appendix discusses a selected set of experiments that have and/or will search for dark photons.
They are organized by dark-photon production method, or the first method used in cases where there are more than one.
Proposals for future searches are only included here if they expect to
produce results
within the next five years.

\subsection{Electron Beam Dumps}

\textbf{NA64:} published results~\cite{Banerjee:2018vgk,Banerjee:2019hmi} and future projections~\cite{Gninenko:2019qiv} for both visible displaced $\aprime \to e^+e^-$ decays
and invisible $\aprime \to \chi\bar{\chi}$ decays.
NA64 uses a 100\,GeV $e^-$ beam  derived from the CERN SPS proton beam incident on an active (calorimeter) target to search for invisible dark photon decays by a missing energy technique.
The $e^-$ momentum is measured by a micromegas tracking station upstream of, and micromegas, GEM, and straw tube tracking downstream of a pair of dipole magnets.
Synchrotron radiation emitted by electrons in the magnetic field is used to tag electrons and strongly suppress pion contamination.
A hermetic shashlik-type EM calorimeter is followed by a hadronic calorimeter to provide good energy containment for SM particle interactions.
In an alternative configuration, an active target is added upstream of the EM calorimeter to enable searches for dark photon decays to $e^+e^-$. \\

\noindent \textbf{E141, E137, E774, KEK, Orsay}: Data from these older electron beam-dump experiments~\cite{Riordan:1987aw,Bjorken:1988as,Bross:1989mp,Konaka:1986cb,Davier:1989wz} have been recast~\cite{Bjorken:2009mm,Andreas:2012mt} to place constraints on visible $\aprime \to e^+e^-$ decays. Each experiment employed an EM calorimeter placed downstream of substantial shielding, with thicknesses in the range 0.12--179\,m. The decay-region lengths were in the range 2--204\,m.

\subsection{Proton Beam Dumps}

\textbf{SpinQuest}: future projections~\cite{Gardner:2015wea,Tsai:2019mtm} for
displaced visible  $\aprime \to \mu^+ \mu^-$ decays.
SpinQuest (E1039), the successor to the SeaQuest experiment, will use Drell-Yan dimuon pairs produced in polarized targets to study the intrinsic spin of the nucleus. It uses the 120\,GeV main injector proton beam at Fermilab. Hadrons are suppressed by an absorber and beam dump located between the target and the spectrometer. The spectrometer uses multiple tracking stages, with two magnets and an additional absorber before the final tracking stage. The addition of two scintillator hodoscopes enables triggers on muon pairs from the decay of dark photons produced in the beam dump.
The experiment will collect data for two years.  Following this run, an electromagnetic calorimeter will be added to allow searches for dark photons in the $e^+e^-$ final state. This configuration of the detector will be known as DarkQuest, and will start data taking in 2023~\cite{Alekhin:2015byh}. \\

\noindent \textbf{CHARM, NOMAD, PS191, $\nu$-CAL I}: Data from these older proton beam-dump experiments~\cite{Blumlein:1990ay,Blumlein:1991xh,Bergsma:1985qz,Astier:2001ck,Bernardi:1985ny} have been recast~\cite{Blumlein:2011mv,Blumlein:2013cua,Gninenko:2012eq,Gninenko:2011uv}  to place constraints on visible $\aprime \to e^+e^-$ decays. Each experiment employed an EM calorimeter placed downstream of substantial shielding, with thicknesses in the range 64--835\,m. The decay-region lengths were in the range 7--23\,m. \\

\noindent \textbf{LSND}: published search for direct detection of $\chi$ particles~\cite{deNiverville:2011it}.
The LSND experiment, which was located at Los Alamos and ran from 1993 to 1998, produced a huge sample of $\pi^0$ decays by impinging an 800\,MeV  proton beam onto a fixed target, which was either water or a high-$Z$ material~\cite{Aguilar:2001ty}.
The detector consisted of a tank filled with 167 tons of mineral oil doped with organic scintillator material, and an array of 1220 photomultiplier tubes that detected the Cherenkov radiation emitted when particles scattered in the tank. \\

\noindent \textbf{MiniBooNE}: published search for direct detection of $\chi$ particles~\cite{Aguilar-Arevalo:2018wea}.
The MiniBooNE experiment was located in the 8\,GeV BNB proton beam line at Fermilab. After the completion of its original neutrino physics program, the experiment undertook a special one-year run (2013--2014) in which the beam bypassed the neutrino target to directly strike its beam dump. This configuration strongly suppressed the $\nu$ flux in the MiniBooNE detector 490\,m downstream, enabling a search for dark matter produced in $\aprime \to \chi\bar{\chi}$ decays, where the dark photons would have been created in the beam dump. MiniBooNE consisted of 818 tons of mineral oil in a 12.2\,m-diameter tank, optically divided into an inner signal volume and an outer veto region, each viewed by separate sets of photomultiplier tubes. \\

\noindent \textbf{COHERENT}: projections for direct detection of $\chi$ particles.
COHERENT is a set of detectors located 20--30~meters from the Spallation Neutron Source at Oak Ridge National Laboratory, in a low-neutron flux area dubbed {\em neutrino alley}.
The high power 1\,GeV proton beam strikes a mercury target to produce a large number of neutrinos through charged pion decay at rest. The six detector technologies include a 22\,kg  liquid argon scintillation detector, whose performance is the basis of their dark matter studies, and a sodium-doped CsI crystal, which provided the first observation of coherent elastic neutrino-nucleus scattering\,\cite{Akimov:2017ade}.

\subsection{$e^+e^-$ Colliders}

\textbf{KLOE}: published results for prompt visible \aprime decays to the $e^+e^-$, $\mu^+ \mu^-$, and $\pi^+\pi^-$ final states~\cite{Anastasi:2015qla,Anastasi:2018azp}; and future projections for invisible $\aprime \to \chi\bar{\chi}$ decays.
The KLOE experiment is located at the DA$\Phi$NE $e^+e^-$ collider at the Frascati National Laboratory. DA$\Phi$NE operates at the $\phi$ meson resonance, enabling precision studies of the neutral kaon system. The original detector configuration, which collected 2.5\,fb$^{-1}$ between 2001--2006, included a large drift chamber and a lead/scintillating-fiber EM calorimeter within a 0.5\,T solenoidal magnetic field. KLOE-2, which collected an additional 5.5\,fb$^{-1}$ between 2014 and 2018, added a cylindrical GEM vertex tracker. KLOE-2 included a trigger for single photon events, which will enable  searching for invisible dark photon decays. \\

\noindent \textbf{BaBar}:
published searches for prompt visible decays to the  $e^+e^-$ and $\mu^+ \mu^-$ final states~\cite{Lees:2014xha}, and for invisible $\aprime \to \chi\bar{\chi}$ decays~\cite{Lees:2017lec}.
The primary focus of the BaBar experiment was the study of CP violation in the $B$ meson system. It was located at the PEP-II asymmetric $e^+e^-$ collider at SLAC, and collected 500\,fb$^{-1}$ of data at and near the $\Upsilon(4S)$ resonance between 1999 and 2008.
It was a general purpose collider detector, with a silicon strip vertex tracker, a drift chamber, charged hadron particle identification using the DIRC technique, a CsI(Tl) crystal EM calorimeter, and a muon-identification system.
The BaBar search for visible \aprime decays used the full data set. The search for invisible decays, however, used only the 2008 data, approximately 10\% of the total, as this was the only data set that included the necessary single-photon trigger. \\

\noindent \textbf{Belle II}:
future projections for prompt visible decays to the  $e^+e^-$ and $\mu^+ \mu^-$ final states~\cite{Kou:2018nap}, and for invisible $\aprime \to \chi\bar{\chi}$ decays~\cite{10.1093/ptep/ptz106}.
Belle~II is located at the SuperKEKB asymmetric $e^+e^-$ collider at the KEK laboratory, which operates at energies near the $\Upsilon(4S)$.
Using the nanobeam technique, the target peak luminosity is $40\times$ larger than the original KEKB collider.
The primary physics goal of Belle~II is to search for new physics through a wide range of measurements that are sensitive to the presence of heavy virtual particles, and that can be precisely predicted in the SM.
These could include CP violation and other asymmetries, rare decays, or forbidden decays. Belle~II will also search for the direct production of new light particles. Data taking started in 2019, and is expected to continue through 2026. \\

\noindent \textbf{LEP}: published results from the DELPHI experiment~\cite{Abdallah:2003np,Abdallah:2008aa} were recast as limits on invisible \aprime decays in Ref.~\cite{Fox:2011fx}.
LEP produced $e^+e^-$ collisions from 1989--2000. When it first started operating, it ran at the $Z$ pole, but after a few upgrades it eventually reached $\sqrt{s} = 209$\,GeV.
DELPHI was a general-purpose detector and one of the four main LEP detectors.
Its mono-photon data set was used to place limits on invisible \aprime decays.

\subsection{LHC Experiments}

\noindent \textbf{LHCb}: published results~\cite{LHCb-PAPER-2017-038,Aaij:2019bvg} and future projections~\cite{Ilten:2016tkc} for both prompt and displaced visible $\aprime \to \mu^+\mu^-$ decays,
along with future projections for both prompt and displaced $\aprime \to e^+e^-$ decays, where the \aprime is produced in the decays of charm mesons~\cite{Ilten:2015hya}.
LHCb is a general purpose detector in the forward region, located at the LHC at CERN. It studies heavy flavor physics, including CP violation, rare decays, and new phenomena such as lepton universality violation.
LHCb employs real-time calibration, alignment, and physics analysis.
During long shutdown 2 (2018--2021) LHCb is moving to a triggerless readout system, in which the full detector information is read out every LHC collision; a data rate of 40\,Tb/s.
The instantaneous luminosity will increase by a factor of 5, and when combined with the detector upgrades, will greatly increase the sensitivity for many physics channels, including dark photon decays. \\

\noindent\textbf{CMS}: published results for prompt visible $\aprime \to \mu^+ \mu^-$ decays~\cite{Sirunyan:2019wqq}.
CMS is a large, general purpose detector located at the CERN Large Hadron Collider, which features a 4\,T solenoid magnet.
CMS has a broad physics program with particular focus on searches for new physics in high transverse momentum processes. The highlight to date has been the discovery, along with ATLAS experiment, of the Higgs boson. The experiment is preparing a number of upgrades to deal with high luminosity LHC running, scheduled to begin in 2026. \\

\noindent\textbf{FASER}:
future projections~\cite{Ariga:2018uku} for visible displaced \aprime decays to the $e^+e^-$, $\mu^+ \mu^-$, and $\pi^+\pi^-$ final states.
FASER~\cite{Feng:2017uoz} will search for long-lived light particles produced in $pp$ collisions at the LHC. It will be located directly on the beam collision axis 480\,m from the ATLAS interaction point. The detector consists of a scintillator veto, followed by a 1.5\,m decay volume, a 2\,m long spectrometer, and a shashlik-style EM calorimeter.
The spectrometer has three silicon-strip tracking stations, with three 0.6\,T permanent magnet dipoles.
Background rates, as confirmed by 2018 emulsion measurements, will be low, and are dominated by high-momentum muons. Data taking will start in 2022.

\subsection{Meson and Lepton Decay Experiments}

\textbf{NA48/2}:
published results on $\pi^0 \to \gamma \aprime$ decays followed by visible prompt  $\aprime \to e^+e^-$ decays~\cite{Batley:2015lha}.
NA48/2 undertook a wide range of studies of charged kaon properties, including direct CP violation, using data collected in 2003 and 2004 at the CERN SPS.
The experiment used simultaneous $K^+$ and $K^-$ beams produced from a 400\,GeV proton beam.
The NA48/2 apparatus included a 114\,m long decay volume, two sets of drift chambers separated by a dipole magnet, a scintillator hodoscope for timing, a LKr EM calorimeter, a hadronic calorimeter, and a muon system. \\

\noindent \textbf{NA62}:
published results~\cite{CortinaGil:2019nuo} and future projections for invisible \aprime decays, where the dark photons are produced in $\pi^0$ decays;
and future projections for  visible displaced decays to the $e^+e^-$ and $\mu^+ \mu^-$ final states.
NA62 is the successor to NA48/2. Its goal is to reconstruct 80 of the ultra-rare decay $K^+ \to \pi^+ \nu \bar \nu$ with low background to extract a measurement of $|V_{td}|$ and to search for deviations from SM expectations.
Kaons in a 75\,GeV hadron beam are identified using Cherenkov radiation before decaying in flight.
A magnetic spectrometer and ring-imaging Cherenkov detector characterize the decay products. The apparatus includes a muon veto, and hermetic photon detection out to wide angles.
The nominal detector configuration uses tagged $\pi^0$ mesons produced in $K^+ \to \pi^+ \pi^0$ decays to search for invisible dark photon decays. An alternative {\em beam dump mode} configuration, would use the 400\,GeV proton beam incident on an upstream beam dump to search for visible decays of long-lived dark photons. \\

\noindent \textbf{Mu3e}:
future projections for prompt visible  $\aprime \to e^+ e^-$ decays, where the dark photons are produced in muon decays~\cite{Echenard:2014lma}.
Mu3e will search for the lepton-flavor-violating process $\mu^+ \to e^+e^-e^+$ at the Paul Scherrer Institute. Using a beam of $10^8$ $\mu$/s stopping in the target, Mu3e will achieve a single event sensitivity of $2 \times 10^{-15}$ in 350 data-taking days. The detector consists of a thin pixel tracker in a 1\,T magnetic field, with scintillating fibers and tiles giving timing resolution of $\mathcal{O}(100\,{\rm ps})$. Studies are under way for a new high-intensity muon beam line that could improve sensitivity by a factor of 20.

\subsection{Electron Fixed Target Experiments}

\textbf{A1}: published results on prompt visible $\aprime \to e^+e^-$ decays~\cite{Merkel:2014avp}.
The A1 experiment was located at the Mainz Microtron (MAMI). It searched for dark photon decays to $e^+e^-$ produced by electron Bremsstrahlung, with  180--855\,MeV electron beams incident on a thin tantalum target.
The apparatus includes a pair of high-resolution spectrometers consisting of four layers of drift chambers, scintillators for timing, and gas Cherenkov detectors to distinguish electrons from pions.
The spectrometers were set to their minimum opening angle, with kinematic acceptance adjusted by changing the beam energy. \\

\noindent \textbf{APEX}: published results and future projections for prompt visible $\aprime \to e^+e^-$ decays~\cite{Abrahamyan:2011gv}.
APEX is located in Jefferson Laboratory Hall A. It searches for dark photons produced in electron Bremsstrahlung in a thin tantalum target. Two septum magnets direct the outgoing $e^+$ and $e^-$ into the two arms of the HRS high-resolution spectrometer. Each spectrometer contains two drift chambers, a timing hodoscope, and gas Cherenkov and lead-glass calorimetry for electron identification.
Data with a 2.2\,GeV beam energy were recorded in 2019. Additional runs at 1.1\,GeV, 3.3\,GeV, and 4.4\,GeV are planned. \\

\noindent \textbf{HPS}:
published results~\cite{Adrian:2018scb} and future projections for prompt visible  $\aprime \to e^+e^-$ decays, along with future projections for displaced decays.
HPS searches for dark photons produced by electron Bremsstrahlung on a thin tungsten target~\cite{Moreno:2013mja}. It is located in Jefferson Laboratory Hall~B. The apparatus is a magnetic spectrometer with two sets of silicon microstrip trackers located 0.5\,mm from the beam plane. A lead tungstate scintillating crystal calorimeter provides particle identification and triggering capability. HPS had engineering runs in 2015 at 1.056\,GeV beam energy and in 2016 at 2.3\,GeV.
The first physics run was at 4.56\,GeV in 2019. Additional beam energies are planned for future runs.